\documentclass[conference]{IEEEtran}
\usepackage[T1]{fontenc}
\usepackage{amsmath,amssymb}
\usepackage{graphicx}
\usepackage{booktabs}
\usepackage{xcolor}
\usepackage{tikz}
\usetikzlibrary{arrows.meta,positioning,fit,backgrounds,calc}
\usepackage[hidelinks]{hyperref}
\usepackage{url}
\usepackage{cite}

\newcommand{\addedLut}{1{,}113}
\newcommand{\addedPow}{0.018}

\newcommand{\advMax}{67}
\newcommand{\advMin}{22}
\newcommand{\bramThree}{9.5}
\newcommand{\bramTwoK}{20.5}

\newcommand{\cpuBasePower}{45}
\newcommand{\cpuCores}{12}

\newcommand{\cpuMigrations}{0}
\newcommand{\cpuMinStateBefore}{5}
\newcommand{\cpuModelName}{Intel Core i5-12500H}
\newcommand{\cpuPinned}{2}
\newcommand{\cpuPinnedRuns}{40{,}000}
\newcommand{\cpuReps}{5}
\newcommand{\cpuRestMs}{15.6}
\newcommand{\cpuRunsPer}{1{,}000}
\newcommand{\cpuScaleRef}{26.4}
\newcommand{\cpuSteerMs}{17.4}
\newcommand{\cpuThreads}{16}
\newcommand{\cpuTurbo}{4.5}
\newcommand{\cpuTurboPower}{95}
\newcommand{\cvAllAllMed}{2.84}
\newcommand{\cvAllOneMed}{5.98}

\newcommand{\cvTenEightyAll}{20.97}
\newcommand{\edgeLinesMax}{11.14}
\newcommand{\edgeLinesMin}{11.02}

\newcommand{\fps}{30.0}
\newcommand{\frameKbit}{4915.2}
\newcommand{\frameMs}{33.3}

\newcommand{\hwCornerUs}{262.1}

\newcommand{\hwEdgeLines}{11.03}
\newcommand{\hwEdgeUs}{720.6}
\newcommand{\hwSpreadNs}{17}

\newcommand{\hwSteerUs}{274.1}

\newcommand{\lineUs}{65.33}
\newcommand{\loopA}{150}
\newcommand{\loopLambda}{2.0}
\newcommand{\loopLostMm}{100}
\newcommand{\loopPreview}{0.30}
\newcommand{\loopSeeds}{8}
\newcommand{\loopT}{30}
\newcommand{\lutScale}{1.03}

\newcommand{\maxMedMax}{12.5}
\newcommand{\maxMedMin}{1.6}
\newcommand{\mhzTenEighty}{74.3}

\newcommand{\nLatPoints}{26}

\newcommand{\nThreadsAll}{16}
\newcommand{\opencvVersion}{4.5.4}
\newcommand{\pixRatio}{27}

\newcommand{\readFrameMs}{6.1}
\newcommand{\refAllMed}{9.87}

\newcommand{\repSpreadMax}{13.7}

\newcommand{\shareOneCorePct}{18}
\newcommand{\shareRefPct}{30}
\newcommand{\shareTenEightyPct}{63}
\newcommand{\stateAllKbit}{213.8}
\newcommand{\stateAllPct}{4.3}
\newcommand{\steerLineLeft}{left}
\newcommand{\sysBram}{41.5}
\newcommand{\sysBramPct}{83}
\newcommand{\sysDsp}{7}

\newcommand{\sysLut}{2{,}568}
\newcommand{\sysLutPct}{12}
\newcommand{\sysPow}{0.318}
\newcommand{\sysWns}{6.316}
\newcommand{\uncontrolledShiftMax}{55}
\newcommand{\verifMpix}{14.8}
\newcommand{\vmaxCpuBest}{4.75}
\newcommand{\vmaxCpuWorst}{3.50}

\newcommand{\vmaxHw}{10.00}
\newcommand{\vmaxIdeal}{9.75}
\newcommand{\wnsTenEighty}{0.245}

\newcommand{\paperOneEdgeUs}{720.63}

\begin{document}

\title{Hardwired Distributed Processing Units at the Sensor: Truly Parallel Vision Pre-Processing that Frees the Host CPU}

\author{\IEEEauthorblockN{Raj Mali}
\IEEEauthorblockA{\textit{Independent Researcher}\\
{raj.mali.research@gmail.com}\\
ORCID: \href{https://orcid.org/0009-0001-6351-0062}{0009-0001-6351-0062}}}

\maketitle

\begin{abstract}
A vision system that sends every pixel to its host processor makes that processor wait for each complete frame and then spend part of every frame period on routine pre-processing. We propose instead to distribute this work into \emph{hardwired distributed processing units} (HDPUs): small fixed datapaths attached directly to the sensor, each dedicated to one always-on task, all operating truly in parallel on the same pixel stream and passing compact results to the host. From a single library of $3\times3$ operators with a common two-clock latency, we build three HDPUs (Canny edges, Harris corners and a line tracker that emits a steering decision) and run them side by side on one camera stream in a low-cost Artix-7 FPGA, using \sysLut{} LUTs and an estimated \sysPow~W. Every stage of every HDPU is bit-exact with its software model. In a full-system simulation with a datasheet camera model, the HDPUs deliver each decision within \hwEdgeLines{} camera lines of its pixel, \advMin--\advMax$\times$ sooner than the earliest moment a frame-based program on the host could, with a latency that varies by at most \hwSpreadNs~ns, whereas the host's own run times reached \maxMedMax$\times$ their median. Offloading the three tasks returns to the host \shareOneCorePct--\shareRefPct\,\% of a core in every frame period at $640\times480$, and \shareTenEightyPct\,\% of the frame period, using all its threads, at 1080p. HDPU cost grows with frame width only through line buffers, and in a closed-loop model of a line-following vehicle the HDPU raises the highest speed at which the line is held from \vmaxCpuBest{} to \vmaxHw~m/s.
\end{abstract}


\section{Introduction}
In most vision systems, the camera sends every pixel to a central processor, which stores the frame and then runs a program over it. Modern processors do this well: with SIMD units, multiple cores and a library such as OpenCV~\cite{bradski2000}, a laptop processes a VGA frame through several detectors in a few milliseconds~\cite{lee2010}. Yet this organisation makes the processor the bottleneck for two reasons. It cannot decide anything until the whole frame has arrived, which takes tens of milliseconds; and the routine pre-processing (edges, corners, a tracking signal) consumes part of every frame period, every frame, leaving less time for the work that actually needs a processor.

We argue for a distributed organisation in which such routine, time-critical tasks run in \emph{hardwired distributed processing units} (HDPUs) placed at the sensor, in the spirit of near-sensor computing~\cite{zhou2020}. An HDPU is a fixed datapath for one task, with no instruction stream, operating system or frame buffer. Pixels flow from the sensor pins through its logic, and every result leaves a fixed number of clock cycles and image lines after its input. Several HDPUs observe the same pixel stream at the same time, each with its own logic, so they are parallel in the literal sense: no unit waits for, shares a bus with, or is scheduled against another. The host remains the master core: it receives compact results (an edge map, a list of corners, a steering code) instead of raw frames, and spends its time on the flexible work that only it can do. We call the always-on task that an HDPU performs a \emph{reflex}.

A companion paper~\cite{paperone} presented one such unit, a Canny edge detector~\cite{canny1986} built from a library of $3\times3$ operators with a common two-clock latency. This paper asks whether the approach generalises to a set of cooperating units, and what the system gains from them. Our contributions are:
\begin{enumerate}
\item Three HDPUs from one operator library, the Canny unit and two new ones, a Harris corner detector~\cite{harris1988} and a dark-line tracker that turns morphological filtering~\cite{haralick1987} into a steering decision, running in parallel on one camera stream with one shared Gaussian stage in a single low-cost FPGA (Section~\ref{sec:arch}).
\item A verification chain in which every HDPU is bit-exact, stage by stage, across a Python model, RTL simulation, a full-system simulation with a camera model, and a C++ port (Section~\ref{sec:verif}).
\item A quantified case for offloading: decision latency from the sensor, its variation, the share of the host's frame period returned to it, and the data that must be held before a decision, measured against carefully controlled host software (Section~\ref{sec:results}).
\item A resolution-scaling study from $320\times240$ to $1920\times1080$, and a closed-loop model that shows what the HDPU's latency means for a vehicle (Sections~\ref{sec:scale} and~\ref{sec:loop}).
\end{enumerate}

\section{Background}
\textbf{Computing near the sensor.} Moving processing towards the sensor reduces the data that must be transferred and the time to a result~\cite{zhou2020}. Horowitz showed that, in a general-purpose processor, the energy of an arithmetic operation is small compared with that of fetching instructions and moving data~\cite{horowitz2014}, and Kuon and Rose quantified the cost of programmable logic relative to custom circuits~\cite{kuon2007}. An FPGA datapath sits between the two: its clock is slower than a processor's, but it fetches no instructions and its structure follows the algorithm, so its effort goes into the computation itself.

\textbf{Streaming vision on FPGAs.} Local image operators map naturally onto line-buffered FPGA pipelines~\cite{neoh2004}. Stencil compilers such as Darkroom~\cite{hegarty2014} and vendor libraries such as Vitis Vision~\cite{vitisvision} generate such pipelines, and multi-resolution architectures combine edge and corner detection on one device~\cite{possa2014}. Most of this work reports throughput and resources; we focus instead on what a set of such units gives the system around them: the time from sensor to decision, its variation, and the load removed from the host.

\textbf{Specialisation and offload.} Hameed et al.\ found a general-purpose chip multiprocessor 500$\times$ less energy efficient than an ASIC on H.264 encoding, with about 90\,\% of its energy spent on overhead rather than arithmetic~\cite{hameed2010}. Conservation cores place small specialised datapaths beside a processor to cut the energy of frequently run code~\cite{venkatesh2010}, and datacenters offload routine work from host CPUs to FPGAs, for search ranking~\cite{putnam2014} and for networking~\cite{firestone2018}. LogCA shows that such offload pays only when the cost of invoking the accelerator and moving data to it is small relative to the work~\cite{altaf2017}. HDPUs follow the same reasoning but sit on the sensor stream itself: the host never invokes them or copies data to them; it only receives their results.

\textbf{Fair baselines.} Lee et al.\ showed that many reported accelerator speed-ups vanish once the processor's code is optimised~\cite{lee2010}. We therefore measure the host running OpenCV, single- and multi-threaded, and a bit-exact C++ port of the HDPUs' algorithms, under controlled conditions.

\textbf{Delay in feedback loops.} In sampled-data control, a delay between measurement and actuation erodes the phase margin and limits achievable performance~\cite{astrom1997}. Our closed-loop experiment uses a pure-pursuit path tracker~\cite{coulter1992} to make this effect concrete.

\section{Three Parallel HDPUs from One Operator Library}\label{sec:arch}
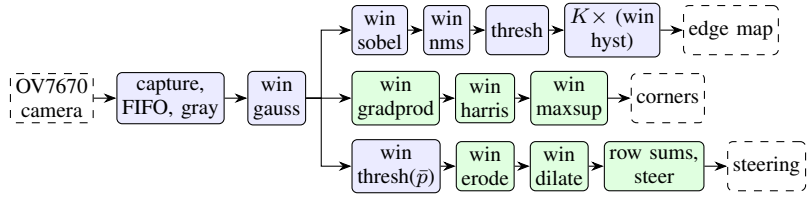
\begin{figure*}[t]
\centering
\begin{tikzpicture}[font=\footnotesize, >=Stealth,
  b/.style={draw, rounded corners=2pt, minimum height=6.5mm, align=center, inner sep=2pt},
  op/.style={b, fill=blue!8},
  newop/.style={b, fill=green!12},
  io/.style={b, dashed}]
\node[io] (cam) {OV7670\\camera};
\node[op, right=3mm of cam] (cap) {capture,\\FIFO, gray};
\node[op, right=3mm of cap] (g) {win\\gauss};
\node[op, right=6mm of g, yshift=9mm] (s) {win\\sobel};
\node[op, right=2mm of s] (n) {win\\nms};
\node[op, right=2mm of n] (t) {thresh};
\node[op, right=2mm of t] (h) {$K\times$ (win\\hyst)};
\node[io, right=3mm of h] (eo) {edge map};
\node[newop, right=6mm of g] (gp) {win\\gradprod};
\node[newop, right=2mm of gp] (hr) {win\\harris};
\node[newop, right=2mm of hr] (ms) {win\\maxsup};
\node[io, right=3mm of ms] (co) {corners};
\node[op, right=6mm of g, yshift=-9mm] (dk) {win\\thresh($\bar{p}$)};
\node[newop, right=2mm of dk] (er) {win\\erode};
\node[newop, right=2mm of er] (di) {win\\dilate};
\node[b, right=2mm of di, fill=green!12] (rs) {row sums,\\steer};
\node[io, right=3mm of rs] (so) {steering};
\draw[->] (cam) -- (cap); \draw[->] (cap) -- (g);
\draw[->] (g.east) -- ++(2mm,0) |- (s.west);
\draw[->] (g) -- (gp);
\draw[->] (g.east) -- ++(2mm,0) |- (dk.west);
\foreach \a/\c in {s/n,n/t,t/h,h/eo,gp/hr,hr/ms,ms/co,dk/er,er/di,di/rs,rs/so} \draw[->] (\a) -- (\c);
\end{tikzpicture}
\caption{Three HDPUs working in parallel on one camera stream: Canny edges (top), Harris corners (middle) and line tracking (bottom). Every box marked ``win'' is a line-buffered $3\times3$ window; every operator produces its result exactly two clock cycles after its window. Blue: operators of the companion paper~\cite{paperone}; green: operators added here. Only the Gaussian stage is shared; the Canny unit uses $K=8$ local hysteresis stages.}
\label{fig:arch}
\end{figure*}

\subsection{Operator contract}
Every operator takes a $3\times3$ window and produces its result exactly two clock cycles later; validity, coordinates and the image-border flag travel alongside in matched delay registers outside the operator. A streaming window holds two image lines in one block RAM and presents a new window for every input sample, so a chain of $S$ windows decides a pixel about $S$ image lines after it arrives. Each stage forces its one-pixel frame border to zero, exactly as the software models do. Because all operators share one timing contract, a new HDPU is a new chain of windows and operators rather than a new control design.

\subsection{Parallel by construction}
The parallelism of the HDPUs is structural rather than scheduled. Every operator in every unit accepts a new pixel in every clock cycle, so in any given cycle each operator of all three units is working on a different pixel. The units share only the Gaussian output; they have no common bus, arbiter, cache or scheduler, and none of them can delay another. A unit's latency is therefore set by its own chain of operators alone, and adding a unit costs logic but no time. The measurements confirm this: in the companion paper, the Canny unit alone decides an edge \paperOneEdgeUs~$\mu$s after its pixel~\cite{paperone}; with the Harris and line units running beside it, the same decision follows its pixel by \hwEdgeUs~$\mu$s (Section~\ref{sec:results}), identical to the precision reported. This differs from parallelism on a processor, where threads compete for cores, caches and memory bandwidth, and where the operating system decides when each task runs.

\subsection{The three HDPUs}
Fig.~\ref{fig:arch} shows the system, and Table~\ref{tab:ops} lists the operators each unit uses.

\textbf{Canny edges} is the fast path of~\cite{paperone}: Gaussian smoothing, Sobel magnitude and quantised direction, non-maximum suppression, double thresholding and $K=8$ local hysteresis stages. It never reports an edge that full hysteresis would reject.

\textbf{Harris corners}~\cite{harris1988} reuses the Gaussian output. A new operator forms the gradients and the three products $g_x^2$, $g_y^2$ and $g_xg_y$ (scaled to 12, 12 and 13 bits); the next sums them over the window and computes $R=\det M-\tfrac{1}{16}(\operatorname{tr}M)^2$ with three DSP multipliers, saturated to an 8-bit score; a third retains local maxima above a threshold. Ties between equal neighbours are resolved in favour of the first pixel in raster order, so a flat maximum yields exactly one corner.

\textbf{Line tracking} finds a dark line on a light floor, as a line-following robot does. The library's threshold operator, applied to the inverted pixel, marks dark pixels; erosion followed by dilation (an opening~\cite{haralick1987}) removes specks narrower than three pixels. A reducer then accumulates, for every row, the number of line pixels and the sum of their $x$ coordinates. At the end of a chosen look-ahead row, it compares the sum with the count multiplied by $W/2\pm16$, which requires no divider, and outputs a steering code: none, left, straight or right.

Together, the three units turn a 16-bit RGB565 pixel stream into a one-bit edge map, a sparse set of corners and one steering code per frame: pre-processed results that the host can use directly.

\begin{table}[!htb]
\caption{Operators used by each HDPU (all with a two-clock latency)}
\label{tab:ops}
\centering\footnotesize
\setlength{\tabcolsep}{4pt}
\begin{tabular}{@{}llccc@{}}
\toprule
Operator & Function & Canny & Harris & Line \\
\midrule
gauss & $3\times3$ binomial smoothing & \checkmark & shared & shared \\
sobel & gradient magnitude, direction & \checkmark & & \\
nms & non-maximum suppression & \checkmark & & \\
thresh & double threshold & \checkmark & & \checkmark \\
hyst & one hysteresis step & $8\times$ & & \\
gradprod\textsuperscript{\textdagger} & $g_x^2,\ g_y^2,\ g_xg_y$ & & \checkmark & \\
harris\textsuperscript{\textdagger} & window sums, response $R$ & & \checkmark & \\
maxsup\textsuperscript{\textdagger} & $3\times3$ maximum, threshold & & \checkmark & \\
morph\textsuperscript{\textdagger} & erosion / dilation & & & $2\times$ \\
\bottomrule
\multicolumn{5}{@{}l}{\scriptsize \textsuperscript{\textdagger}New in this paper. The others are from~\cite{paperone}.}
\end{tabular}
\end{table}

\subsection{The host}
The host is an \cpuModelName{} with \cpuCores{} cores, \cpuThreads{} threads and a maximum frequency of \cpuTurbo~GHz~\cite{intelark}. To quantify what the HDPUs take off it, we run the same three tasks on the host in two forms. The first is a C++ port that is bit-exact with the HDPUs. The second uses OpenCV \opencvVersion{}: Gaussian blur followed by Canny; the Harris response with dilation-based maximum suppression; and inverse thresholding, morphological opening and row reduction. OpenCV is the code a practitioner would actually write; it is not bit-exact with the HDPUs, nor does it need to be.

\section{Verification and Methodology}\label{sec:verif}
\textbf{Bit-exact chain.} Each HDPU has a Python model that mirrors the hardware operation for operation, including shifts, saturation, border zeroing and tie-breaking. Table~\ref{tab:verif} lists every comparison. The RTL cores are simulated with Icarus Verilog on photographs, an RGB565 frame delivered with camera-like gaps, synthetic line-follower frames and noise, and every intermediate stream is compared with the model. The complete system (capture, clock-domain-crossing FIFO, frame controller, three HDPUs and frame buffers) is simulated in Vivado's simulator against a behavioural OV7670 model that follows the datasheet's VGA timing~\cite{ov7670}, and the cores are additionally simulated at four frame widths. The complete system also runs on the board with a live OV7670 camera and VGA output. In total, \verifMpix{} million output values are compared, and not one differs.

\begin{table}[!htb]
\caption{Verification against the bit-exact models}
\label{tab:verif}
\centering\footnotesize
\begin{tabular}{@{}p{0.37\columnwidth}p{0.33\columnwidth}rr@{}}
\toprule
Comparison & Compared & Runs & Differ \\
\midrule
Canny core (RTL) vs.\ model & every stage, $K=1,2,4,8$ & 20 & 0 \\
Harris and line cores (RTL) vs.\ models & every stage, row sums, steering & 7 & 0 \\
Cores at four frame widths (RTL) vs.\ models & edges, corners, line mask & 4 & 0 \\
Three-HDPU system with camera model vs.\ models & edges, corners, line mask, steering & 2 & 0 \\
C++ port vs.\ models & all outputs, four resolutions & 11 & 0 \\
\bottomrule
\end{tabular}

\end{table}

\textbf{HDPU measurements.} Latency is measured in the system simulation, from the camera edge that delivers a pixel to the clock edge at which its decision leaves the unit, for edges and corners at six positions across the frame and for the steering decision from the last pixel of the look-ahead row. All designs are placed and routed with Vivado 2024.2 for the XC7A35T-1 at 50~MHz; single units and the scaling study are implemented out of context.

\textbf{Controlled host measurements.} In preliminary, uncontrolled runs on the same laptop, medians shifted by up to \uncontrolledShiftMax\,\% between two sessions, so we control the conditions. The benchmark refuses to start while other heavy tools are running or on battery power; it raises the minimum processor state of the power plan from \cpuMinStateBefore\,\% to 100\,\% for the run and restores it afterwards; it pins single-thread programs to one performance core (logical CPU \cpuPinned) and records the core on which each run ends; and it repeats every program in \cpuReps{} independent processes of \cpuRunsPer{} timed runs after warm-up, shuffling the order of programs in every repetition. Programs run under the Windows Subsystem for Linux and are compiled with \texttt{g++ -O3 -march=native}. Timing covers the processing call only, so camera transfer and operating-system wake-up are excluded; every host latency reported below is therefore a lower bound, and every host load an underestimate.

\textbf{Derived quantities.} The host's sensor-to-decision latency is derived as the time from a pixel's arrival to the end of the active frame, computed from the camera timing (\lineUs~$\mu$s per line, \fps{} frames per second), plus the measured median processing time. The share of the frame period returned to the host is the measured median processing time divided by the frame period (\frameMs~ms). Such quantities are labelled \emph{derived}.

\section{Results}\label{sec:results}
\subsection{Cost of the HDPUs}
Table~\ref{tab:cost} gives the post-route cost. The complete system, comprising camera capture, three HDPUs, two frame buffers and VGA output, uses \sysLut{} LUTs (\sysLutPct\,\% of the device), \sysBram{} block RAMs (\sysBramPct\,\%) and \sysDsp{} DSPs, and closes timing at 50~MHz with \sysWns~ns of slack. Adding the Harris and line units to the Canny-only system costs \addedLut{} LUTs and, by Vivado's estimate, \addedPow~W. Most of the block RAM holds the two display frame buffers, not the HDPUs.

\begin{table}[!htb]
\caption{Hardware cost on the XC7A35T-1 (Vivado 2024.2)}
\label{tab:cost}
\centering\footnotesize
\resizebox{\columnwidth}{!}{\begin{tabular}{@{}lrrrrrr@{}}
\toprule
Design & LUT & FF & BRAM\textsuperscript{a} & DSP & WNS & Power\textsuperscript{b} \\
 & & & & & (ns) & (W) \\
\midrule
Canny core, $K=8$ & 848 & 1{,}768 & 6.0 & 2 & 8.047 & 0.080 \\
Harris core & 720 & 844 & 3.5 & 3 & 5.634 & 0.091 \\
Line-tracker core & 375 & 525 & 1.5 & 2 & 6.923 & 0.076 \\
\midrule
Complete system\textsuperscript{c} & 2{,}568 & 3{,}472 & 41.5 & 7 & 6.316 & 0.318 \\
Device XC7A35T & 20{,}800 & 41{,}600 & 50 & 90 & & \\
\bottomrule
\multicolumn{7}{@{}p{0.97\columnwidth}@{}}{\scriptsize All at 50~MHz, post-route. \textsuperscript{a}36-kbit blocks. \textsuperscript{b}Vivado estimate, default activity; cores out of context, without I/O. \textsuperscript{c}Camera capture, FIFO, the three HDPUs, two frame buffers and VGA output.}
\end{tabular}
}
\end{table}

\subsection{The load the HDPUs remove from the host}
Table~\ref{tab:cpu} gives the host's time for the three tasks under controlled conditions. On a stored $640\times480$ frame, OpenCV needs \cvAllAllMed~ms with \nThreadsAll{} threads and \cvAllOneMed~ms on a single core, and the bit-exact C++ port \refAllMed~ms. The host is fast; a pipeline that accepts one pixel per clock at 50~MHz needs \readFrameMs~ms merely to read a stored frame (derived), so for images already in memory the host remains the right tool. At the camera's \fps{} frames per second, however, this time is spent in every frame period: \shareOneCorePct\,\% of one core with OpenCV and \shareRefPct\,\% with the C++ port (derived). At $1920\times1080$, OpenCV takes \cvTenEightyAll~ms using all \nThreadsAll{} threads, \shareTenEightyPct\,\% of the frame period (derived). With the HDPUs in place, this time returns to the host, which receives finished results instead. The host's run times are also uneven: under controlled conditions, the medians of the \cpuReps{} repetitions agree to within \repSpreadMax\,\%, and pinned runs ended on their assigned core in all but \cpuMigrations{} of \cpuPinnedRuns{} cases, yet in every program the slowest run took \maxMedMin--\maxMedMax$\times$ the median.

\begin{table}[!htb]
\caption{Host time per stored frame under controlled conditions: overall median, range of the per-repetition medians, 99.9th percentile, and worst run relative to the median}
\label{tab:cpu}
\centering\footnotesize
\resizebox{\columnwidth}{!}{\begin{tabular}{@{}lrrrr@{}}
\toprule
Program, $640\times480$ (ms) & median & per rep. & p99.9 & max/med \\
\midrule
\multicolumn{5}{@{}l}{\emph{C++ port (bit-exact), 1 thread, pinned}} \\
\quad Canny fast path, $K=8$ & 6.58 & 6.52--6.63 & 9.85 & 2.1 \\
\quad Harris corners & 4.21 & 4.16--4.31 & 7.29 & 2.4 \\
\quad Line tracker & 0.85 & 0.84--0.86 & 1.75 & 12.5 \\
\quad All three & 9.87 & 9.76--10.01 & 16.22 & 3.1 \\
\quad Exact Canny, in-place & 5.81 & 5.80--5.83 & 9.86 & 2.0 \\
\quad Exact Canny, two-buffer & 13.86 & 13.73--13.98 & 21.12 & 2.0 \\
\multicolumn{5}{@{}l}{\emph{OpenCV, 1 thread, pinned}} \\
\quad Canny & 3.59 & 3.52--3.69 & 5.67 & 2.5 \\
\quad Harris corners & 2.19 & 1.93--2.23 & 4.02 & 2.4 \\
\quad Line tracker & 1.78 & 1.76--1.80 & 3.50 & 5.1 \\
\quad All three & 5.98 & 5.87--6.07 & 10.18 & 2.0 \\
\multicolumn{5}{@{}l}{\emph{OpenCV, 16 threads}} \\
\quad Canny & 0.52 & 0.51--0.54 & 0.86 & 1.7 \\
\quad Harris corners & 1.84 & 1.84--1.87 & 3.37 & 2.6 \\
\quad Line tracker & 1.82 & 1.81--1.84 & 3.55 & 3.0 \\
\quad All three & 2.84 & 2.75--3.13 & 4.35 & 1.6 \\
\bottomrule
\end{tabular}
}
\end{table}

\subsection{Decisions while the frame is still arriving}
\begin{figure}[!htb]
\centering
\includegraphics[width=\columnwidth]{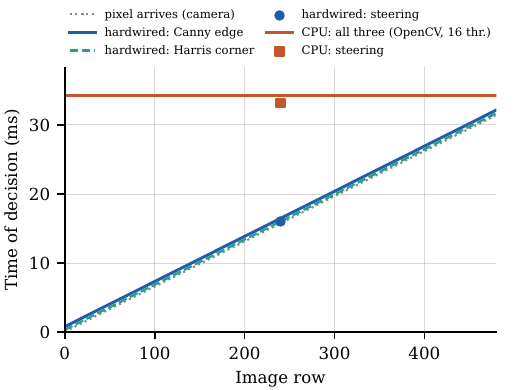}
\caption{Time at which decisions are available, from the start of the active frame. HDPUs (hardwired): system simulation with the camera model. Host (CPU): derived from the camera timing and the measured median processing time (lower bound).}
\label{fig:timeline}
\end{figure}

Fig.~\ref{fig:timeline} shows when each decision becomes available during a frame. The HDPUs track the camera at a fixed distance: an edge decision follows its pixel by \hwEdgeUs~$\mu$s (\hwEdgeLines{} lines), a corner decision by \hwCornerUs~$\mu$s, and the steering decision arrives \hwSteerUs~$\mu$s after the last pixel of the look-ahead row. A frame-based program on the host can produce any decision only after the last active line has arrived and the frame has been processed. Table~\ref{tab:latency} compares the two: the HDPUs decide \advMin--\advMax$\times$ sooner. For the steering decision of row 240, the host must still wait for the remaining 239 rows, \cpuSteerMs~ms in total even with OpenCV's fastest time. This comparison is with frame-based software; a CPU or microcontroller program written to process the image line by line could also start early, and what the HDPUs add is a latency fixed by their structure and a host that is free of the work. These host figures assume that the frame reaches the processor the instant the camera completes it and that the process is already running; both assumptions favour the host.

\begin{table}[!htb]
\caption{Sensor-to-decision latency. Hardwired HDPUs: system simulation (median over positions and frames). $^{*}$Host CPU: derived lower bound with OpenCV, \nThreadsAll{} threads.}
\label{tab:latency}
\centering\footnotesize
\begin{tabular}{@{}lrrrr@{}}
\toprule
 & \multicolumn{2}{c}{Hardwired} & CPU\textsuperscript{*} & Hardwired \\
\cmidrule(lr){2-3}\cmidrule(lr){4-4}
HDPU and decision & $\mu$s & lines & ms & sooner \\
\midrule
Canny: edge pixel & 720.6 & 11.03 & 16.2 & 22$\times$ \\
Harris: corner pixel & 262.1 & 4.01 & 17.5 & 67$\times$ \\
Line: steering, row 240 & 274.1 & 4.20 & 17.4 & 64$\times$ \\
\bottomrule
\end{tabular}

\end{table}

\subsection{Determinism}
\begin{figure}[!htb]
\centering
\includegraphics[width=\columnwidth]{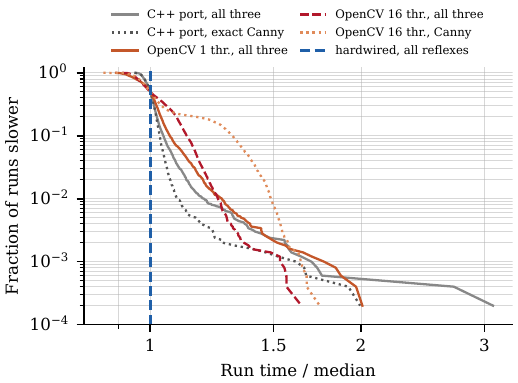}
\caption{Fraction of host runs slower than a given multiple of the median (controlled benchmark, all repetitions pooled). Dashed vertical line: largest relative deviation of any HDPU latency.}
\label{fig:tails}
\end{figure}

\begin{figure}[!htb]
\centering
\includegraphics[width=\columnwidth]{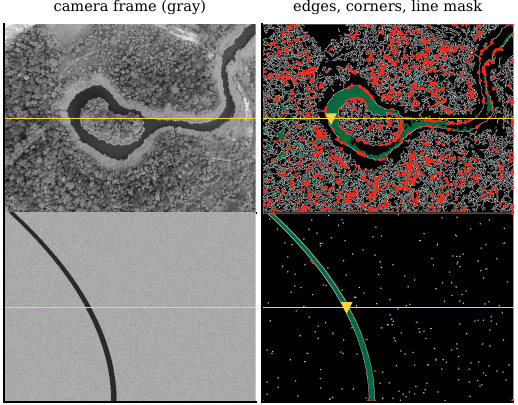}
\caption{Outputs of the three HDPUs (models, bit-exact with the RTL) for an aerial photograph of a river and a synthetic line-follower view: Canny edges (white), Harris corners (red), line mask after opening (green), look-ahead row 240 (yellow) and the line centroid on it. In the photograph, the line unit isolates the dark river channel.}
\label{fig:example}
\end{figure}

Fig.~\ref{fig:tails} shows the distribution of the host's run times relative to each program's median. Every program has a tail, and for a system with a deadline it is the tail, not the median, that sets the margin. The HDPU latencies, measured at \nLatPoints{} points (two frames, six positions, three units), vary by at most \hwSpreadNs~ns, less than one 20-ns clock cycle; this residue stems from the asynchronous crossing from the camera clock domain. Nothing in an HDPU depends on the image content, on other tasks or on history, so its latency is a property of the design, not of the workload. Fig.~\ref{fig:example} shows the outputs of all three units on an aerial river scene and on a line-follower view; in the latter, the line unit steers \steerLineLeft{} for the curving line, in agreement with the model.

\subsection{Data held before a decision}
A frame-based program needs the whole frame, \frameKbit~kbit in RGB565, before its first decision. The three HDPUs together hold only their line buffers: \stateAllKbit~kbit, or \stateAllPct\,\% of a frame (derived from the window widths). This is the structural reason for the latency gap: an HDPU never holds more of the image than its $3\times3$ operators require.

\subsection{Scaling with resolution}\label{sec:scale}
\begin{figure}[!htb]
\centering
\includegraphics[width=\columnwidth]{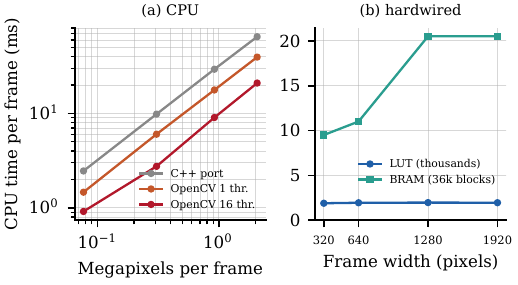}
\caption{(a) Host time per frame for all three tasks grows with the pixel count. (b) HDPU resources for all three units grow only through line buffers.}
\label{fig:scale}
\end{figure}

Table~\ref{tab:scale} and Fig.~\ref{fig:scale} repeat the measurements from $320\times240$ to $1920\times1080$. On the host, time grows with the number of pixels: the C++ port takes \cpuScaleRef$\times$ longer for a frame with \pixRatio$\times$ the pixels. In the HDPUs, the logic barely changes: from the smallest to the largest frame, the LUT count grows by a factor of only \lutScale{}, because only the line buffers and the coordinate counters depend on the width; block RAM grows from \bramThree{} to \bramTwoK. The latency in image lines is constant (\edgeLinesMin--\edgeLinesMax{} lines for an edge at every width), and the fraction of the frame that must be held shrinks as frames grow. At $1920\times1080$, the design also closes timing at \mhzTenEighty~MHz (slack \wnsTenEighty~ns), the pixel clock of 1080p at 30 frames per second, so one pixel per clock suffices even at that resolution. The larger the frame, the more host time the HDPUs save and the less, relatively, they cost.

\begin{table}[!htb]
\caption{Scaling with frame size: HDPUs (out-of-context implementation; RTL simulation for latency) and host (controlled benchmark)}
\label{tab:scale}
\centering\footnotesize
\resizebox{\columnwidth}{!}{\begin{tabular}{@{}lrrrr@{}}
\toprule
Frame & $320{\times}240$ & $640{\times}480$ & $1280{\times}720$ & $1920{\times}1080$ \\
\midrule
\multicolumn{5}{@{}l}{\emph{Hardwired, all three HDPUs (Vivado; RTL simulation)}} \\
\quad LUT & 1{,}889 & 1{,}943 & 1{,}961 & 1{,}949 \\
\quad BRAM (36k) & 9.5 & 11.0 & 20.5 & 20.5 \\
\quad WNS at 50~MHz (ns) & 6.161 & 5.978 & 6.347 & 6.162 \\
\quad Edge latency (lines) & 11.14 & 11.07 & 11.04 & 11.02 \\
\quad State held (\% frame) & 8.7 & 4.3 & 2.9 & 1.9 \\
\multicolumn{5}{@{}l}{\emph{CPU, all three tasks, median ms per frame}} \\
\quad C++ port, 1 thread & 2.46 & 9.84 & 29.39 & 64.85 \\
\quad OpenCV, 1 thread & 1.47 & 6.03 & 17.73 & 39.49 \\
\quad OpenCV, 16 threads & 0.92 & 2.75 & 9.05 & 20.97 \\
\bottomrule
\end{tabular}
}
\end{table}

\subsection{Does latency matter in a loop?}\label{sec:loop}
To assess what the latency means for a machine, we model a line-following vehicle (model; details in the repository). A unicycle drives at constant speed along a sinusoidal line (amplitude \loopA~mm, wavelength \loopLambda~m). Once per frame, the camera's look-ahead row 240 observes the line \loopPreview~m ahead, and a pure-pursuit rule~\cite{coulter1992} sets the curvature from that observation. All cases see the same rows of the same frames; only the time at which the decision is applied differs. For the HDPU, this is the measured steering latency. For the host, it is the remainder of the frame, \cpuRestMs~ms, plus a processing time drawn from that program's measured distribution, which again omits transfer and wake-up. The line counts as lost when the lateral error exceeds \loopLostMm~mm, and each speed is run \loopSeeds{} times for \loopT~s.

\begin{figure}[!htb]
\centering
\includegraphics[width=\columnwidth]{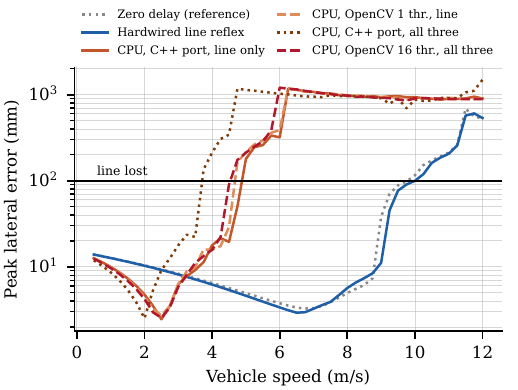}
\caption{Peak lateral error of the line-following vehicle versus speed, for different sources of the steering decision (model, with measured latencies).}
\label{fig:loop}
\end{figure}

Fig.~\ref{fig:loop} and Table~\ref{tab:loop} show the result. With the HDPU, the vehicle holds the line up to \vmaxHw~m/s, practically the same as with zero delay (\vmaxIdeal~m/s; the two differ by one step of the speed sweep, which we do not interpret as an advantage of the delay). With the same decision computed on the host from the same rows, the limit falls to \vmaxCpuWorst--\vmaxCpuBest~m/s, depending on the program. The exact figures depend on the vehicle and controller, but the mechanism is general~\cite{astrom1997}: the delay of frame-based processing is a large fraction of the frame period, and it is incurred in every control cycle.

\begin{table}[!htb]
\caption{Highest speed at which the line is held in every run (model). $^{*}$Rest of the frame plus median processing time; the model draws processing times from the measured distribution.}
\label{tab:loop}
\centering\footnotesize
\begin{tabular}{@{}lrr@{}}
\toprule
Steering decision from & delay (ms) & max.\ speed (m/s) \\
\midrule
Zero delay (reference) & 0.00 & 9.75 \\
Hardwired line reflex & 0.27 & 10.00 \\
CPU, C++ port, line only & 16.46$^{*}$ & 4.75 \\
CPU, OpenCV 1 thr., line & 17.39$^{*}$ & 4.50 \\
CPU, C++ port, all three & 25.48$^{*}$ & 3.50 \\
CPU, OpenCV 16 thr., all three & 18.46$^{*}$ & 4.50 \\
\bottomrule
\end{tabular}

\end{table}

\subsection{Power}
Vivado estimates the complete three-HDPU system at \sysPow~W, of which the Harris and line units add \addedPow~W. The host's documented base power is \cpuBasePower~W, with a turbo limit of \cpuTurboPower~W~\cite{intelark}. The two figures are not a like-for-like comparison (a tool estimate for the HDPUs, a datasheet value for the host), so we draw no ratio from them; the difference is nevertheless consistent with the instruction and data-movement overheads described by Horowitz~\cite{horowitz2014}, which an HDPU does not incur. A host relieved of the pre-processing can also spend more of its time in low-power states.

\section{Discussion}
\textbf{Why distribute the work.} The measurements show what each part of the system does best. The host is fast and flexible on data in memory. The HDPUs decide while the frame is still arriving, with a latency that does not vary, while holding a few lines rather than a frame and drawing, by Vivado's estimate, a fraction of a watt. Their advantage comes not from clock speed (50~MHz against \cpuTurbo~GHz) but from organisation: each unit is dedicated to one task, works on the pixel stream as it arrives, and runs in parallel with every other unit without competing for shared resources.

\textbf{HDPUs and a host.} The two are complementary. A set of hardwired HDPUs next to the sensor handles the always-on, time-critical pre-processing and passes compact results (edge maps, corner lists, a steering code) to the host, which performs the deliberate and flexible work with the time the HDPUs free. Because every HDPU is built on the same operator contract and verified against a model, adding one is a matter of composing operators, as the Harris and line units demonstrate, and it adds no latency to the units already present.

\textbf{Limitations.} The hardware results derive from RTL and system simulation with a datasheet camera model and from Vivado implementation; the complete system runs on the board with a live camera, but its latency was measured in simulation, and measuring latency and power on the board remains to be reported. The host figures cover one laptop processor under the Windows Subsystem for Linux; its latencies are derived lower bounds and its freed load is derived from processing times. The closed-loop experiment is a model with an idealised vehicle and a simple controller; it demonstrates the effect of the measured latencies, not a validated vehicle.

\section{Conclusion}
A single library of $3\times3$ operators with a common two-clock latency yields three hardwired distributed processing units (Canny edges, Harris corners and a steering line tracker) that run truly in parallel on one camera stream in a low-cost FPGA and are bit-exact with their software models. Placed at the sensor, the HDPUs decide \advMin--\advMax$\times$ sooner than a frame-based program on the host could, with a latency that varies by at most \hwSpreadNs~ns, while holding only \stateAllPct\,\% of a frame and drawing an estimated \sysPow~W. They return to the host \shareOneCorePct--\shareRefPct\,\% of a core in every frame period at VGA resolution, and their cost grows with resolution only through line buffers. In a closed-loop model, they raise the speed at which a vehicle can follow a line from \vmaxCpuBest{} to \vmaxHw~m/s. These results support distributing routine, time-critical vision into hardwired units at the sensor, and reserving the host processor for the work that needs its flexibility.

\bibliographystyle{IEEEtran}
\bibliography{refs}

\end{document}